**Complementary use of TEM and APT for the investigation of steels nanostructured by severe plastic deformation.**

X. Sauvage[1], W. Lefebvre[1], C. Genevois[1], S. Ohsaki[2], K. Hono[2]

1- University of Rouen, Groupe de Physique des Matériaux, CNRS (UMR 6634), F-76801 Saint-Etienne du Rouvray, France

2- National Institute for Materials Science, 1-2-1, Sengen, Tsukuba, 305-0047, Japan

**Abstract**

The properties of bulk nanostructured materials are often controlled by atomic scale features like segregation along defects or composition gradients. Here we discuss about the complimentary use of TEM and APT to obtain a full description of nanostructures. The advantages and limitations of both techniques are highlighted on the basis of experimental data collected in severely deformed steels with a special emphasis on carbon spatial distribution.

Corresponding author:
Xavier Sauvage: xavier.sauvage@univ-rouen.fr

**Introduction**

Strength of materials has been known to increase as inverse proportional to the crystal grain size following the Hall-Petch relationship. Thus, continuous efforts have been made to improve the strength of structural materials by refining crystal grain sizes by controlling recrystallization by thermomechanical treatments. In steels and aluminum alloys, fine grain size means 1 to 10 μm, and there have not been effective processing technique to reduce the grain size of these materials to less than 100 nm until recently. Although mechanical milling and alloying can process the powders containing nanosized grains, grain growth cannot be suppressed during consolidation processes like sintering and hot pressing. Therefore, processing bulk nanostructured materials (BNM) for structural applications is still a big challenge, in particular using an industrially viable process. Recent developments of various severe plastic deformation (SPD) processes [1, 2] like equal channel angular pressing (ECAP), accumulative roll bonding (ARB), and high pressure torsion (HPT) straining opened up a new way to process nanosized grain microstructures to steels and aluminum alloys, and it is in principle capable of scaling up ECAP and ARB to an industrial scale. The main advantage of these SPD techniques is that materials are free of porosity unlike the powder consolidation processes. The final grain size is typically less than 300 nm in most of metals and alloys which gives rise to a significant strengthening; however the main drawbacks of BNMs are the lack of ductility and of thermal stability [3]. Therefore, various attempts are being made to achieve both high strength and ductility by modifying the nanostructures.

Many studies on SPD have been applied to single phase materials to refine grain size. Applications of SPD to multiphase microstructures have also been attempted to explore the possibility of obtaining ultrafine composite nanostructures, e.g. [4-10]. One classical example of severe plastic deformation of two phase alloys can be seen in well known pearlite wire that is widely used as suspension cables, tire cords and fishing wires. By cold drawing pearlite to a high strain of more than 4, the cementite and ferrite lamellar spacing is reduced to a few tens nanometers, and a strength exceeding 3 GPa can be commonly obtained [11-13]. Since cementite lacks plasticity, they are fragmented in the initial drawing stage and eventually decomposed to carbon and ferrite. A similar concept was used to develop ultrahigh strength electric wires for pulsed magnets by cold drawing Cu-Ag eutectic microstructures. Although Cu and Ag are insoluble each other, they are considered to be mixed each other after large strain deformation like mechanical alloying [14]. To observe such decomposition and mixing phenomena that occur during severe plastic deformation of multiphase alloys, analytical techniques that have a rather high spatial resolution and quantitative analytical capabilities are required. Transmission electron microscopy (TEM) with energy dispersive X-ray spectroscopy (EDS) as well as electron energy loss spectroscopy (EELS) is the most versatile technique for qualitative analysis of nanostructures processed by SPD, but atom probe tomography (APT) has superior spatial resolution as well as quantitative analytical capability of embedded particles and light elements. Effective complimentary applications of TEM and APT will provide critical information regarding the mechanism of nanostructure evolution of multi-phase alloys processed by SPD. In this article, we critically overview the current problems as to the nanostructures of multiphase alloys processed by SPD, using as example the typical case of steels.

**APT and TEM**

APT is a projection type field ion microscope (FIM) combined with a time of flight mass spectrometer [15]. Atoms are ionized from a hemispherical surface of a sharp needle specimen (FIM tip) of approximately 50 nm in radius by the field evaporation phenomenon. Positively charged ions accelerated along the electric field are detected on a position sensitive detector located in front of the specimen. Since atoms are field-evaporated layer-by-layer, APT has an atomic layer resolution in the depth direction of the elemental map with optimized analysis conditions. The mass resolution depends on the energy deficit of ions as well as the flight length, but is usually improved by energy compensator like reflectrons [16]. The area for analysis depends on the acceptance angle $2\theta$ of the detector which is determined by the tip radius, tip detector distance and the diameter of the detector ($\theta \sim tan^{-1}(D/d)$, $AA \sim 2\pi(r\theta)^2$, where $D$: diameter of detector, $d$: tip detector distance, $AA$: analysis area, $r$: tip radius). Because of recent wide angle configuration of APT, typical size of analysis volumes has expanded from 10x10x200 $nm^3$ in the 90's to 50x50x200 $nm^3$ nowadays, meaning that the number of collected atoms has expanded from typically half a million to several tens of millions. Although it is possible to achieve an atomic resolution in the depth direction, the lateral resolution is subject to the evaporation aberration, which is about 0.3 nm. Nevertheless, the APT is the only analytical microscope that can provide three dimensional real elemental maps of constituent elements in alloys.

During the last twenty years APT has been used widely to measure the composition of nanoscale particles in various metallic materials and to display their distribution and morphology. It is also a powerful instrument to reveal chemical gradients or to highlight segregation along structural defects like grain boundaries or dislocations. Recently, it has been demonstrated that femtosecond laser could successfully assist the field evaporation of materials analyzed by APT [17]. It reduces the stress applied to the specimen, thereby reducing the probability of specimen failure. This feature is extremely important for the atom probe analysis of nanostructured materials that are prone to rupture during voltage pulse field evaporation due to the presence of a high number density of defects like dislocations and grain boundaries. APT specimens have been prepared for a long time by standard electropolishing techniques but nowadays micro-milling by the focused ion beam (FIB) technique is widely used [18], which made site specific specimen preparation possible, for example from surface layers, grain boundaries and powders [19].

Compared to APT, TEM provides images of micro- or nano-structures with a larger field of view, and chemical mapping can easily be achieved by energy filtering TEM (EFTEM) or scanning TEM (STEM) using EDS, high angle annular dark field (HAADF) or EELS. Moreover, crystallographic information can be collected using electron diffraction techniques or high resolution TEM (HRTEM) images. Nevertheless, one should note that both EDS and EELS are rather qualitative. This makes difficult the measurement of concentrations typically lower than 1at.% in nanoscaled grains or the detection of segregation along crystallographic defects like grain boundaries. Moreover, the composition of nanosized particles embedded in the matrix cannot be quantitatively measured. Although TEM tomography was recently developed to build up the three-dimensional tomography of microstructures, most of the case they give only morphological information without any chemical information [20].

**Complimentary TEM and APT analysis of cold drawn pearlitic steel**

Cold drawn pearlitic steel wires are among the strongest commercial steels (yield stress up to 3GPa) and are widely used in the tire industry [11-13]. This extremely high strength results from their unique nanostructure that is progressively formed during the cold drawing process: cementite lamellae are aligned along the wire axis, and the interlamellar spacing is reduced in a range of 10 to 30 nm, while the thickness of the lamellae is typically less than 5 nm [11-13]. At the end of the drawing process, the typical cumulated true strain applied to the material is about 3 for commercial products and higher than 4 for laboratory material. It was often reported that such high level deformation also leads to the decomposition of cementite [21-27]. This feature was first discovered by Mössbauer spectroscopy and then confirmed by APT measurements in the late 90's. However, the physical mechanisms leading to this decomposition are still under debate [9, 12, 27-29], especially because the description of the resulting nanostructure and of the carbon distribution is not fully understood. In some cases, carbon supersaturated solutions in the α-Fe matrix exhibiting some features similar to quenched martensite have even been reported [22, 23, 30].

The driving force and the kinetics of this phase transformation are still controversial. Some authors point out the possible effect of moving dislocations that cross cementite lamellae. They may trap carbon atoms of the lamellae and then redistribute them inside the ferrite phase [21]. Others argue that the driving force for the decomposition could be the dramatic increase of the $Fe_3C$/α-Fe interfacial area resulting from the lamellae thinning. This feature may lead to a strong increase of the carbon solubility in the ferrite through the well known Gibbs-Thomson effect [12, 27]. Cementite decomposition is still a challenging issue since it strongly affects the deformation mechanisms (dislocation nucleation and motion) in the drawn wires, and thus its mechanical properties and especially the ductility and the ageing behaviour. In spite of the large amount of experimental data published in the literature, the influence of metallurgical parameters (original interlamellar spacing, pearlite colony size, alloying elements) and of processing parameters is not yet well established. In the following, we demonstrate that linking HR-STEM data to advanced APT data treatment lead to a better understanding of such unique nanostructure.

The nanostructure of drawn pearlitic steels is usually imaged using bright field TEM, but as shown in Fig.1, it can also be observed by FIM (Fig. 1(a)) or HAADF STEM (Fig. 1(b)). The dim contrast of cementite in the FIM image results from the concavity of the specimen surface due to the lower evaporation field of cementite compared to that of the α-Fe matrix. When the image is observed at a lower temperature like 20K, these contrasts disappear. This means that no information on the concentration of carbon in each phase can be obtained from the FIM images. On the other hand, HAADF images are sensitive to the atomic number, so the darkly imaged regions in Fig. 1 (b) are considered to contain higher amount of carbon compared to the matrix. Thus, in the HAADF image, the contrast of the recorded images can be linked to the carbon concentration gradient within the nanostructure. This indicates that even if cementite is partly decomposed as it was pointed out by Mössbauer spectroscopy long time ago [21], a lamellar structure is kept and carbon atoms are not homogeneously distributed. Using APT, it is possible to obtain the three-dimensional carbon map of such a lamella (Fig. 2(a)). These data confirm that most of carbon atoms are located in thin lamellae of cementite or former cementite with a thickness in a range of 2 to 4 nm. However, as seen in the concentration profile (Fig. 2 (c)) that was calculated across the lamella of the volume displayed in the Fig. 2 (a), the carbon concentration in the α-Fe phase is in a range of 0.5 to 1 at.%, i.e. much higher than the equilibrium solubility limit. Note that the maximum carbon concentration is well below the 25at.% for the stoechiometry of $Fe_3C$, confirming the strain induced decomposition of this carbide. However, due to the lower evaporation field of cementite, local magnification effects and ion trajectory aberrations could significantly affect the distribution of carbon atoms within the reconstructed volume [31]. Thus, the large carbon gradients exhibited on each side of the interface by the composition profile (Fig. 2(c)) may not reflect the actual concentration of carbon.

To check this point, EELS mapping of a carbon rich region with a similar interlamellar spacing was performed (Fig. 3(a)). The profile of the intensity of the carbon K edge (Fig. 3(b)) was computed (with the three windows method) across the lamella with a sampling volume similar to the one used for the carbon concentration profile of the APT data (Fig. 2(c)). Due to carbon contamination on the TEM sample surface, the carbon concentration cannot be accurately measured in the $\alpha$-Fe phase. Despite this, the profile unambiguously shows that there are large carbon gradients in the $\alpha$-Fe phase. The gradients spread over a distance of about 2 nm in each side of the interface, in accordance with the APT data. Thus, we can conclude that local magnification effects do not significantly affect the distribution of carbon atoms within the APT reconstructed volume. From the carbon EELS map (Figure 3(a)), it is clear that the carbon concentration is not homogeneous within the probed lamella (there are less bright dots on the left). A careful treatment of the APT data can confirm this point too. The original data set displayed in the Fig. 2(a) was filtered to highlight in grey the regions containing more than 10at.% carbon (see details of the procedure in [32]). Some clusters with a diameter in range of 2 to 5 nm do appear within the carbon rich lamellae. They could be some remaining cementite nano-particles. This inhomogeneous concentration fluctuation of carbon in the former cementite lamellae is considered to be due to the fragmentation of the cementite lamellae into nanocrystalline cementite as was observed by the dark field TEM image in [25].

Therefore, HR-STEM was carried out to image the nanostructure at the atomic scale in a region where the interlamellar spacing is only about 10 nm (Fig.4). Carbon rich lamellae are darkly imaged on the HAADF image (arrowed in the Fig. 4(b)). Due to lattice distortions, the bright field STEM image (Fig. 4(a)) is difficult to interpret directly thus it was Fourier filtered (Fig. 4(c)). Nine dislocations are arrowed (in white), but due to distortions all of them are probably not visible in this image. Thus, in this region, the dislocation density is at least 2 $10^{16}m^{-2}$. One should note that dislocations are not exclusively located in the regions with the highest carbon concentration (dark zones in the HAADF image). Thus, one can conclude that carbon atoms resulting from the cementite decomposition are not homogeneously distributed on interstitial sites in the bcc lattice of the α-Fe phase but obviously trapped by stress fields of dislocations. This confirms the critical role of dislocations in the decomposition process of cementite [9, 21, 28].

This can be further supported by the APT results of carbon distribution in both mechanically milled pearlite and mechanically milled martensite. As reported by Ohsaki [33], cementite in pearlite powders are completely decomposed into carbon, resulted in the formation of nanocrystalline ferrite. Figure 5 shows (a) TEM, (b) FIM, and (c) APT results of Fe-3.8at.%C pearlite powder that was mechanically milled for 100 h. The TEM image shows complete decomposition of cementite. The ring pattern observed inside the {110} ring is due to the formation of oxide, and it does not correspond to cementite. The FIM image shows nanocrystalline features of ferrite, i.e., the brightly imaged regions correspond to ferrite and the darkly imaged regions correspond to the grain boundaries. The contrast mechanism is similar to that for Fig. 1 (a). The APT data show that the carbon distribution is not uniform regardless of the complete decomposition of cementite; carbon atoms are segregated along the grain boundaries. Within the ferrite, a small amount of C is detected, but majority of carbon are detected from the grain boundaries. This suggests that carbon tend to be segregated at defects. When nanocrystalline ferrite is formed, most of dislocations are swept away from the nanocrystals to form high angle grain boundaries; hence the carbon atoms segregated at dislocations are also dragged to the grain boundaries. This scenario can be further supported by the carbon distribution in nanocrystalline ferrite that was formed by mechanical milling of martensite powder. Figure 6 shows (a) TEM, (b) FIM and (c) APT results obtained from the nanocrystalline ferrite powder that was processed by mechanically milling Fe-3.6at.%C martensite steel powder [34]. Since it was martensite before mechanical milling, we assume that C distribution was uniform in the unmilled condition. Although the nanostructure formation does not involve $Fe_3C$ decomposition, after long time milling that is enough to produce nanocrystalline ferrite, the microstructure observed by TEM, FIM and APT are nearly the same as was produced from the mechanically milled pearlite. This suggests that the final microstructures were nearly the same; i.e., most carbon atoms are segregated along the grain boundaries with a small amount of carbon in the ferrite. Since the carbon distribution was uniform in the original martensite structure, the carbon depletion inside ferrite nanocrystals strongly suggests that it was swept away by dislocations to form high angle grain boundaries. Since there is little chance for a dislocation can be present within the nanocrystalline ferrite, carbon dissolution in the nanocrystalline ferrite is considered to be low.

**Concluding remarks**

TEM is without any doubt a more versatile characterization technique than APT, especially because it provides both structural and chemical information. Moreover, recent developments in analytical TEM (STEM, EELS, EDS) have led to chemical analysis down to the atomic column scale. However, there are still numerous applications where the complimentary use of APT is effective, especially for nanocrystalline alloys, in particular when low concentrations of light elements play a critical role in nanostructuring mechanisms or stability. Indeed, only this technique is able to quantitatively analyze nanoscaled features that are embedded in the matrix phase and smaller than the thin foil thickness. In the present paper, we highlighted a unique nanostructure feature of cold drawn pearlitic steels, and demonstrated that the complementary characterization by TEM and APT is useful. The solute segregations along structural defects is also a typical example where the APT has a superior analytical ability. It is especially relevant for nanostructured alloys with an inherent high volume fraction of grain boundaries and it is also a critical point for nanostructured materials processed by SPD with high dislocation densities. However, APT reconstructions are not always free from artifacts like local magnification effects, which may affect particle shapes and related composition gradients. As shown in this paper, analytical STEM may help to clarify doubts. It points out once more the evident complimentarily of APT and TEM that may continue to grow in the future, especially in the field of nanostructured alloys. Today, the main drawback is that TEM foils and APT needle shape samples are very different, thus one cannot analyze in optimal conditions exactly the same feature with both microscopes. However, recent instrumental developments like FIB nano-machining [18] or TEM specimen holder designed for tips [35] may push ahead the complimentary use of APT and TEM, especially in the field of tomography at the nanometer scale [36].

**Acknowledgements**

Dr Eiji Okunishi from JEOL Ltd is gratefully acknowledged by the authors for the TEM observations. The authors thank Mr. Guillaume Lathus and Yuichi Matsumoto from JEOL Ltd for providing access to the JEOL 2100F TEM at Akishima (Japan). This work was in part supported by the Grant-in-Aid for Scientific Research on Priority Areas ''Giant Straining Process for Advanced Materials Containing Ultra-High Density Lattice Defects'', MEXT, Japan, the World Premier International Research Center Initiative (WPI Initiative) on Materials Nanoarchitronics, MEXT, and CREST, JST.

**Figure captions**

**Figure 1** Cold drawn pearlitic steel (Fe-3.6at.%C, true strain 3.5) (a) FIM image showing the nanostructure of the cold drawn pearlitic steel. Cementite lamellae (arrowed) have a dark contrast due to their lower evaporation field comparing to the $\alpha$-Fe matrix. (b) STEM HAADF image of the same nanostructure. Cementite lamellae (arrowed) have as well a dark contrast because the atomic number of C (Z=6) is much smaller than that of Fe (Z=26).

**Figure 2** Cold drawn pearlitic steel (Fe-3.6at.%C, true strain 3.5) (a) High resolution STEM BF image of the nanostructure with FFT (inset). (b) STEM HAADF image of the same region where dark lamellae corresponding to cementite are arrowed. (c) Fourier filtered image of the BF. Some dislocations are arrowed in white.

**Figure 3** Cold drawn pearlitic steel (Fe-3.6at.%C, true strain 3.5) (a) Carbon K edge STEM EELS map, bright dots correspond to high carbon concentration. (b) EELS profile computed across the carbon rich lamella exhibited on the EELS map (obtained in a section of 5 nm as imaged on the EELS map).

**Figure 4** Cold drawn pearlitic steel (Fe-3.6at.%C, true strain 3.5) (a) 3D reconstructed volume of a small volume analyzed by APT. Only carbon atoms are plotted to exhibit a carbon rich lamella. (b) Same volume with regions containing more than 10at.% highlighted in grey. (c) Composition profile computed across the lamella (averaged across a section of $5 \times 5 nm^2$ as imaged on the reconstructed volume and with a sampling volume thickness of 1 nm)

**Figure 5** (a) TEM, (b) FIM, and (c) APT map of the nanocrystalline ferrite produced from the Fe-3.6at.%C pearlite steel wire by mechanically milling for 100 h. Data reproduced from reference [33]

**Figure 6** (a) TEM, (b) FIM, and (c) APT map of the nanocrystalline ferrite produced from the Fe-3.6at.%C martensite steel wire by mechanically milling for 100 h. Reproduced from reference [34]

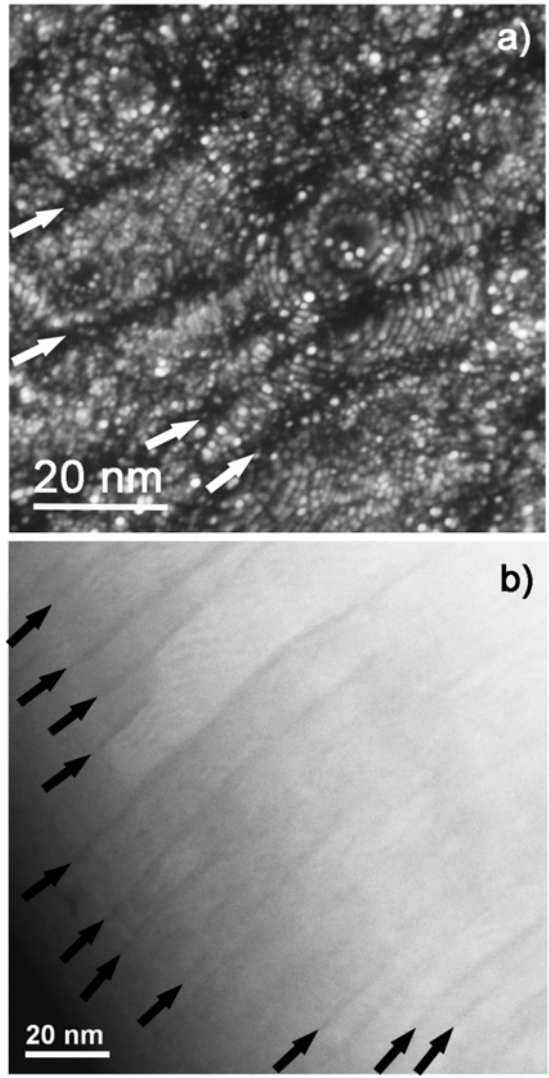

Figure 1

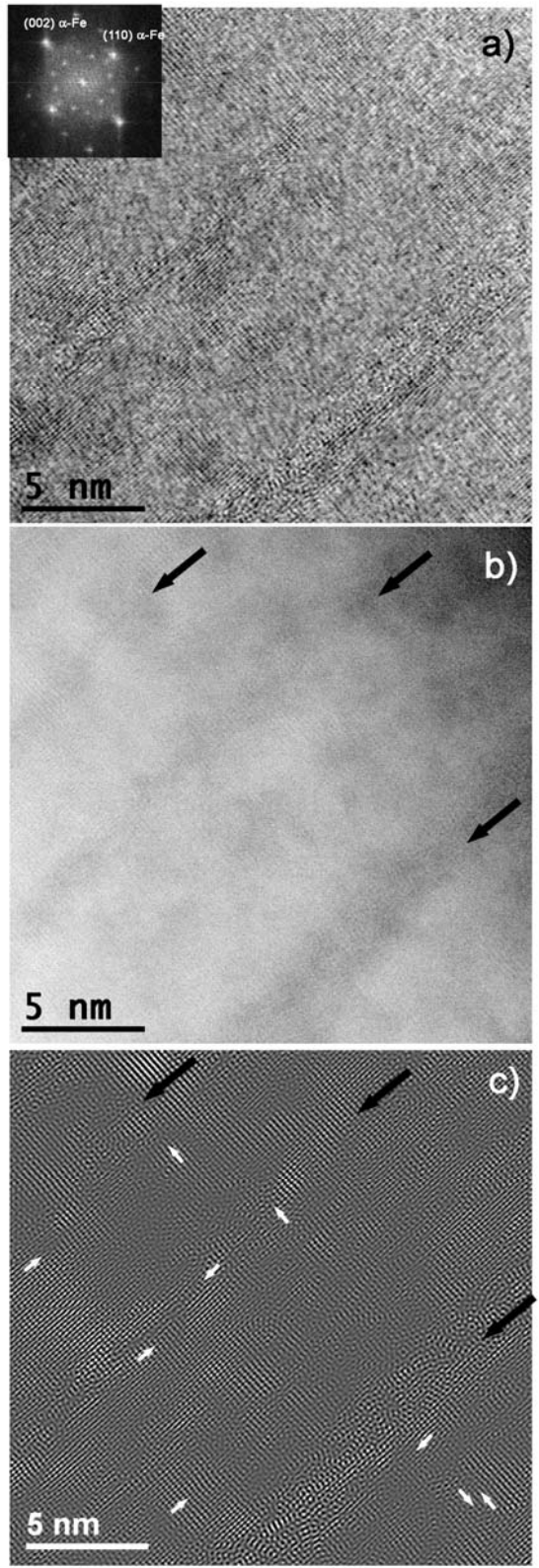

Figure 2

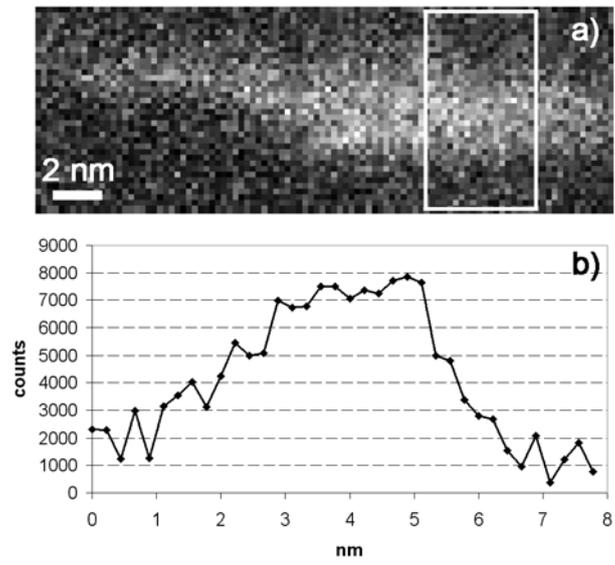

Figure 3

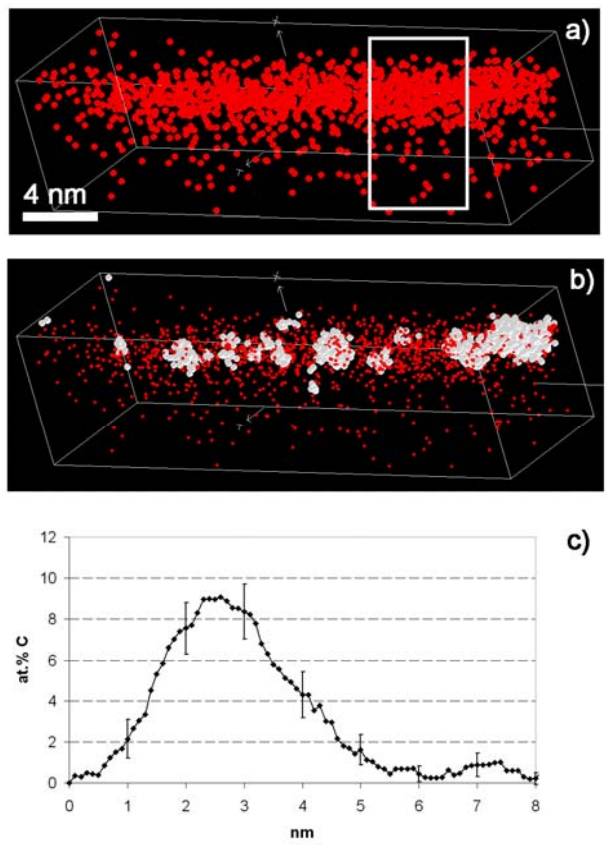

Figure 4

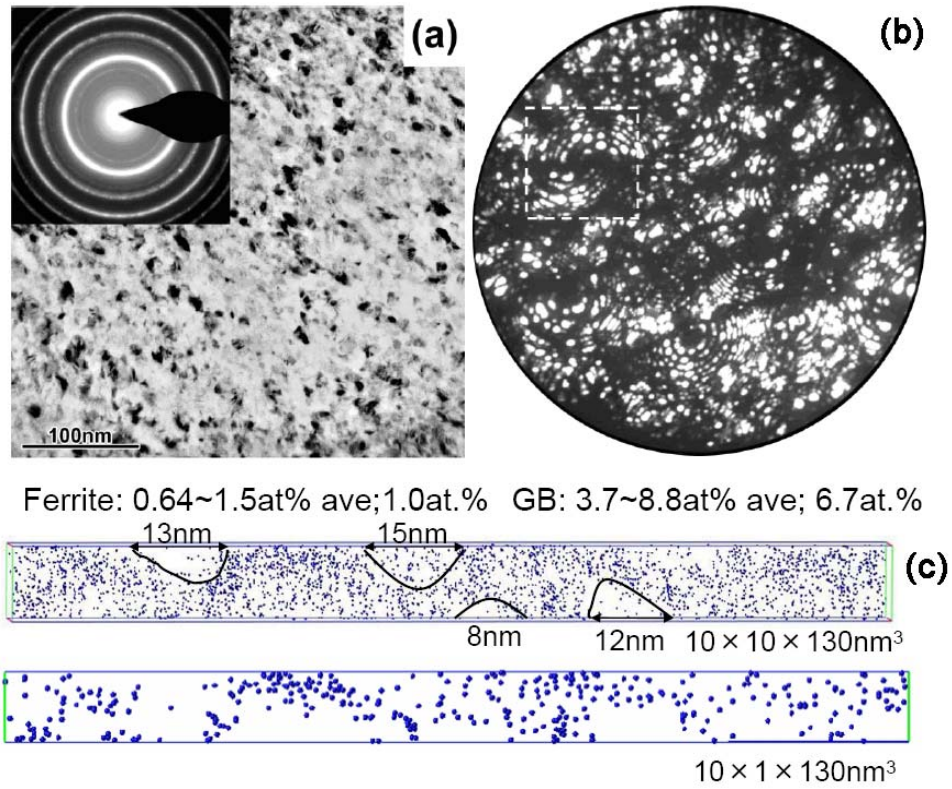

Figure 5

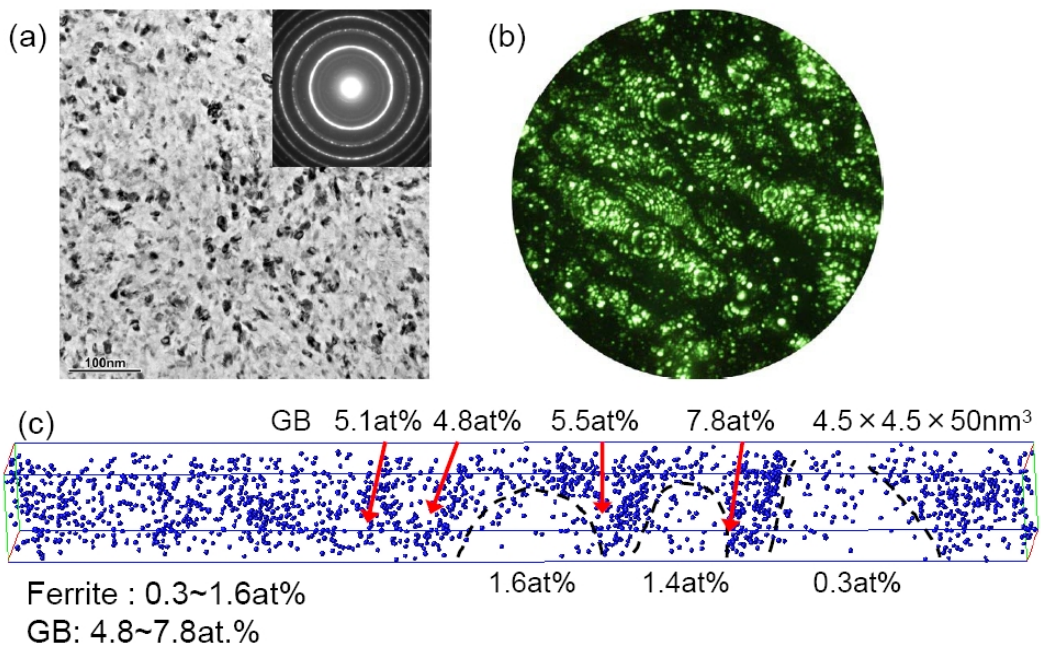

Figure 6